\documentclass[a4paper,fleqn,usenatbib]{mnras}

\usepackage{newtxtext,newtxmath}

\usepackage[T1]{fontenc}
\usepackage{ae,aecompl}
\usepackage{lipsum} 

\usepackage{graphicx}	
\usepackage{amsmath}	
\usepackage{amssymb}	

\title[Dynamical evolution of triple-star systems]{Dynamical evolution of triple-star systems by Lidov-Kozai cycles and tidal friction}

\author[M. Bataille et al.]{
M. Bataille$^{1}$,
A.-S. Libert$^{1}$\thanks{E-mail: anne-sophie.libert@unamur.be},
A. C. M. Correia$^{2,3,4}$
\\
$^{1}$naXys, Department of Mathematics, University of Namur, Rempart de la Vierge 8, B-5000 Namur, Belgium \\
$^{2}$Department of Physics, University of Coimbra, 3004-516 Coimbra, Portugal\\
$^{3}$CIDMA, Department of Physics, University of Aveiro, 3810-193 Aveiro, Portugal\\
$^{4}$ASD, IMCCE, Paris Observatory, PSL University, 77 Av. Denfert-Rochereau, 75014 Paris, France\\
}

\date{Accepted XXX. Received YYY; in original form ZZZ}

\pubyear{2018}

\begin{document}
\label{firstpage}
\pagerange{\pageref{firstpage}--\pageref{lastpage}}
\maketitle

\begin{abstract}
Many triple-star systems have an inner pair with an orbital period of a few days only. A common mechanism to explain the short-period pile-up present in the observations is the migration through Lidov-Kozai cycles combined with tidal friction. Here, we revisit this mechanism and aim to determine the initial orbital configurations leading to this process. We show that the mutual inclination of the triple-star system is not the only critical parameter, since the eccentricity as well as the argument of the pericenter of the inner orbit also play an important role in the establishment of the Lidov-Kozai migration. Our framework is the secular hierarchical three-body problem (octupole order approximation) with general relativity corrections, including the effects of tides, stellar oblateness and magnetic spin-down braking. Both the orbital and the spin evolutions are considered. Extensive numerical simulations with uniform and non-uniform distributions of the initial orbital parameters are carried out, and unbiased initial conditions leading to Lidov-Kozai migration are revealed. Finally, we highlight the importance of the initial ``Kozai constant'' $h=\sqrt{(1-e^2)}\cos{i}$ in the dynamical evolution of triple-star systems, by showing that phase portraits at given $h$-values unveil different evolution paths. 
\end{abstract}

\begin{keywords}
binaries (including multiple): close -- celestial mechanics -- stars: kinematics and dynamics -- methods: analytical -- methods: numerical

\end{keywords}



\section{Introduction}\label{sec:Introduction}

Although the statistics of multiple star systems are quite incomplete and suffer from discovery biases, many observational studies have reported a significant proportion of triple-star systems with an inner binary period smaller than six days (e.g., \citealt{Duquennoy_Mayor_1991, Tokovinin_1997, Tokovinin_2001, Tokovinin_2014}). A possible well-accepted mechanism for the observed pile-up around three days is the Lidov-Kozai mechanism \citep{Lidov_1962, Kozai_1962} combined with tidal friction (e.g., \citealt{Eggleton_Kiseleva-Eggleton_2001, Fabrycky_Tremaine_2007, Naoz_Fabrycky_2014, Liu_et_al_2015, Toonen_et_al_2016, Anderson_et_al_2017}). For an inner binary perturbed by an outer inclined stellar companion (tertiary), the eccentricity of the inner orbit can reach values close to unity during the Lidov-Kozai cycles, for which the periastron distance can be very small. Tidal dissipation is very efficient at this point and causes the inner orbit to shrink. This mechanism is often referred to as {\it Lidov-Kozai migration}. It has also been widely invoked for the formation of Hot Jupiters in binary systems (e.g., \citealt{Wu_Murray_2003, Naoz_et_al_2012, Petrovich_2015_jan, Anderson_et_al_2016_mar}). 

Concerning triple-star systems, \cite{Fabrycky_Tremaine_2007} have studied the secular evolution of hierarchical triple-star systems using the quadrupole approximation with general relativity and tidal effects. An over-abundance of short-period binaries result from their Monte-Carlo simulations, in agreement with the observations, and could be interpreted as a possible indication of the presence of a distant tertiary companion in short-period binaries. They have also noticed that tertiary companions leading to Lidov-Kozai migration have preferably a final mutual inclination around $\sim 40^\circ$ or $\sim 140^\circ$ with respect to the inner binary plane. \cite{Naoz_Fabrycky_2014} have extended this study using the octupole approximation. In particular, they have shown that a larger range of initial mutual inclinations leads to the Lidov-Kozai migration mechanism. Further refinements have been proposed by \cite{Anderson_et_al_2017}  and \cite{Moe_Kratter_2017}.

In the present study, we aim to determine more precisely which initial values of the orbital elements initiate the Lidov-Kozai migration, and the subsequent formation of the short-period pile-up. In particular, we show that this kind of migration is not only dependent on the initial mutual inclination of the systems. 
In line with the work of \cite{Fabrycky_Tremaine_2007}, extensive numerical simulations are carried out, using the vectorial secular equations of \cite{Correia_et_al_2016} for the spin and orbital evolution (octupole approximation) with general relativity corrections, including the effects of tides, stellar oblateness and magnetic spin-down braking. 
This model is described in detail in Section~\ref{sec:model}, and the results of the simulations are presented in Section~\ref{sec:nonuniform}. In order to remove any possible biases in the initial distributions of \cite{Fabrycky_Tremaine_2007}, uniform distributions of the orbital parameters are introduced in Section~\ref{sec:uniform}, and unbiaised initial conditions leading to Lidov-Kozai migration revealed. A dynamical analysis of several evolutions is shown in Section~\ref{sec:analytical}, where phase portraits allow to disentangle different evolution paths. Initial conditions for migrating stars are finally discussed in Section~\ref{sec:h} and our results summarized in Section~\ref{sec:ccl}.         

\section{Model}\label{sec:model}
We consider a hierarchical system consisting of an inner binary with masses $m_0$ and $m_1$, and a distant companion star with mass $m_2$. We use Jacobi coordinates $r_j$ $(j=1,2)$: the orbit of $m_1$ relative to $m_0$ is called the inner orbit, and the outer orbit refers to the one of $m_2$ relative to the center of mass of $m_0$ and $m_1$. Both bodies of the inner binary are considered as oblate ellipsoids rotating about the axis of maximal inertia along the direction of the spins $\hat{\textbf{s}}_j$ with rotation rates $w_j$ ($j=0,1$). The $J_2$ gravity field coefficients are given by ($j=0,1$)
\begin{equation}
J_{2_j}=\frac{k_{2_j}w_j^2 R_j^2}{3Gm_j},
\end{equation}
where $R_j$ is the radius of $m_j$ and $k_{2_j}$ is the second Love number.

We study the orbital and spin evolutions using the vectorial secular equations of \cite{Correia_et_al_2016}, which include general relativity corrections, conservative and dissipative tides and stellar oblateness. More precisely, the octupole level of approximation is used to describe the secular orbital interaction (e.g., \citealt{Ford_et_al_2000}). The evolution of the orbits is tracked by the orbital angular momenta $(j=1,2)$
\begin{equation}
\textbf{G}_j=\beta_j \sqrt{\mu_j a_j (1-e_j^2)} \, \hat{\textbf{k}}_j
\end{equation}
and the Laplace-Runge-Lenz vectors $\textbf{e}_j$ along the major axis in the direction of periapsis with magnitude $e_j$ ($j=1,2$)  
\begin{equation}
\textbf{e}_j = \frac{\dot{\textbf{r}}_j \times \textbf{G}_j}{\beta_j \mu_j} -
\frac{\textbf{r}_j }{r_j} \ ,
\end{equation}
where $\hat{\textbf{k}}_j$ is the unit vector $\textbf{G}_j$, $e_j$ the eccentricity of the orbit, $a_j$ the semi-major axis, 
$\mu_1=G (m_0+m_1)$, $\mu_2=G (m_0+m_1+m_2)$,
$ \beta_1 =  G m_0 m_1 / \mu_1 $, and  $ \beta_2 = \mu_1 m_2 / \mu_2 $. 

The evolution of the spins is followed by tracking the rotational angular momenta $(j=0,1)$
\begin{equation}
\textbf{L}_j = C_j w_j \hat{\textbf{s}}_j,
\end{equation}
where $\hat{\textbf{s}}_j$ is the unit vector $\textbf{L}_j$ and $C_j$ the principal moment of inertia. 

Regarding the dissipation of the mechanical energy of tides in the body's interior, we use a weak friction model with constant time delay $\Delta t$  \citep{Mignard_1979}. The magnetic spin-down for the two stars of the inner binary is taken from \cite{Skumanich_1972} and \cite{Anderson_et_al_2016_mar}, with $\dot{w_j} \propto -w_j^3$. 
Concerning the general relativity, we only consider the pericenter precession of the inner orbit. 

We use as reference plane the initial orbit of $m_2$, thus the initial $\hat{\textbf{k}}_2=(0,0,1)$. 
The initial unit vectors $\hat{\textbf{k}}_1$ and $\textbf{e}_j$ are given by
\begin{eqnarray}
\hat{\textbf{k}}_1&=&R_z{(\omega_1)} \, R_x(i) \hat{\textbf{k}}_2, \\
\textbf{e}_j&=&R_z{(\Omega_j)} \, R_x(i)\, R_z (\omega_j) \, R_y\left(\frac{\pi}{2}\right) \hat{\textbf{k}}_2 \, \, \,\, (j=1,2), 
\end{eqnarray}
where $\omega_j$ is the argument of the pericenter, $\Omega_j$ is the longitude of the ascending node, and $i$ is the mutual inclination between the orbital planes.
The initial $\hat{\textbf{s}}_j$ is given by
\begin{eqnarray}
\hat{\textbf{s}}_j&=&R_z{(\omega_1)} \, R_x(i) R_z{(\phi_j)} \, R_x(\theta_j) \hat{\textbf{k}}_2 \, \, \,\, (j=0,1),
\end{eqnarray}
where $\theta_j$ the obliquity to the orbital plane of the inner orbit, and $\phi_j$ the precession angle on this plane.


\begin{figure}
	\centering \includegraphics[width=1\columnwidth]{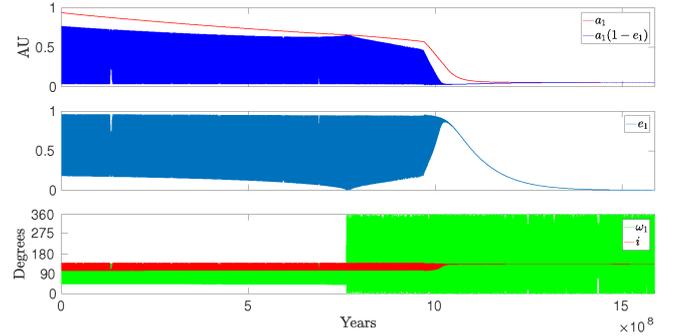}   
	\caption{Example of a triple-star system undergoing Lidov-Kozai migration. Initial conditions of the system are given in Table~\ref{ci_systeme4856}.}
	\label{fig:evolution_systeme4856}
\end{figure}

\begin{table}
\caption{Initial conditions of the Lidov-Kozai migrating system of Fig.~\ref{fig:evolution_systeme4856}.}
\begin{center}
{\scriptsize
\begin{tabular}{|r|r|r|r|r|r|r|}
\hline
 Variables & \multicolumn{2}{r|}{$m_0$} & \multicolumn{2}{r|}{$m_1$} & \multicolumn{2}{r|}{$m_2$} \\
 \hline
 Mass $m_j$ ($m_\odot$) & \multicolumn{2}{r|}{$1$} &  \multicolumn{2}{r|}{$0.84$} &  \multicolumn{2}{r|}{$1.17$} \\
 Radius $R_j$ (m)&  \multicolumn{2}{r|}{$696.34 \times 10^6$} &  \multicolumn{2}{r|}{$604.67 \times 10^6$} &  \multicolumn{2}{r|}{-}\\
 Rotation period $P_{{\rm rot},j}$ (d)&  \multicolumn{2}{r|}{$2.3$} &  \multicolumn{2}{r|}{$2.3$} &  \multicolumn{2}{r|}{-} \\
 Obliquity $\theta_j$ (deg)&  \multicolumn{2}{r|}{$153.95$} &  \multicolumn{2}{r|}{$53.79$} &  \multicolumn{2}{r|}{-} \\
 Precession $\phi_j$ (deg)&  \multicolumn{2}{r|}{$195.35$} &  \multicolumn{2}{r|}{$71.50$} &  \multicolumn{2}{r|}{-} \\
 $C_j$ &  \multicolumn{2}{r|}{$0.08 m_0 R_0^2$} &  \multicolumn{2}{r|}{$0.08 m_1 R_1^2$} &  \multicolumn{2}{r|}{-} \\
 Love number $k_{2_j}$ &  \multicolumn{2}{r|}{$0.028$} &  \multicolumn{2}{r|}{$0.028$} &  \multicolumn{2}{r|}{-} \\
 $\Delta t_j$ (s)&  \multicolumn{2}{r|}{$0.5$} &  \multicolumn{2}{r|}{$0.5$} &  \multicolumn{2}{r|}{-} \\
\hline
 Semi-major axis $a_j$ (AU)& \multicolumn{2}{r|}{} & \multicolumn{2}{r|}{$0.94 $} & \multicolumn{2}{r|}{$72.56$} \\
 Eccentricity $e_j$ & \multicolumn{2}{r|}{} & \multicolumn{2}{r|}{$0.19$} & \multicolumn{2}{r|}{$0.88$} \\
 Argument of pericenter $\omega_j$ (deg)& \multicolumn{2}{r|}{} & \multicolumn{2}{r|}{$101.06$} & \multicolumn{2}{r|}{$144.43$}  \\
 Longitude of the node $\Omega_j$ (deg)& \multicolumn{2}{r|}{} & \multicolumn{2}{r|}{$300.83$} & \multicolumn{2}{r|}{ $17.26$}  \\
\hline
 Mutual inclination $i$ (deg)&\multicolumn{3}{c|}{} &  \multicolumn{3}{c|}{$\qquad 105.32$}    \\
\hline
\end{tabular}
}
\end{center}
\label{ci_systeme4856}
\end{table} 

\normalsize

\footnotesize
\begin{table*}
\caption{Initial conditions of our simulations.}
\begin{center}
{ 
\begin{tabular}{|r|r|r|}
\hline
 Variables &  Non-uniform distributions & Uniform distributions\\
\hline
   Population size $N$ &  100 000 & 100 000 \\
Integration size &  $t \sim U([0.5 \times 10^9 , 10^{10}]) $ &  $t \sim U([0.5 \times 10^9 , 10^{10}]) $\\
   Disruption parameter $f_t$ & 2.46 & 2.46 \\
\hline
 Mass $m_0$ ($m_\odot$) & 1 & 1 \\
   Mass $m_1$ ($m_\odot$) & $\frac{m_1}{m_0} \sim \mathcal{N}(0.23,0.42)$ & $  U([0.1,1.5])$ \\
  Mass $m_2$ ($m_\odot$) &  $\frac{m_2}{m_0+m_1} \sim \mathcal{N}(0.23,0.42)$ & $  U([0.1,1.5])$ \\
\hline
  Orbital period $P_1$ (d) & $\log_{10}(P_1 (d)) \sim \mathcal{N}(4.8,2.3)$ & $\log_{10}(P_1 (d)) \sim U([-2,6])$ \\
  Orbital period $P_2$ (d) & $\log_{10}(P_2 (d)) \sim \mathcal{N}(4.8,2.3)$ & $\log_{10}(P_2 (d)) \sim U([-2,6])$ \\
 \hline
 Eccentricity $e_1$, $e_2$ & $P_j < 1000\, \text{d} \Rightarrow e_j \sim \text{Rayleigh}\,(0.33) $ & $  U([0,1])$ \\
 & $P_j > 1000\, \text{d} \Rightarrow e_j \sim \text{Ambartsumian}$ &   \\
\hline
 Love number $k_{2_0}$ &   0.028  &   0.028 \\
 Love number $k_{2_1}$ &   0.028 &   0.028 \\
\hline
 $\Delta t_0$ ($s$) &  0.5 &  0.5 \\
 $\Delta t_1$ ($s$) &  0.5 &  0.5 \\
 Radius $R_0$ ($R_\odot$) &  1  &  1 \\
 Radius $R_1$ ($R_\odot$) &   $R_\odot \left( \frac{m_1}{m_0} \right)^{0.8} $  &   $R_\odot \left( \frac{m_1}{m_0} \right)^{0.8} $\\
 $C_0$ &   $0.08 m_0 R_0^2$ &   $0.08 m_0 R_0^2$ \\
 $C_1$ &  $0.08 m_1 R_1^2$ &   $0.08 m_1 R_1^2$ \\    
\hline 
 Mutual inclination $i$ &  $  \cos i \sim U([-1,1])$ &  $  \cos i \sim U([-1,1])$\\
 $P_{{\rm rot},0}$ (d) &  2.3&  2.3 \\
 $P_{{\rm rot},1}$ (d) & 2.3 &  2.3\\  
\hline
 Obliquity $\theta_0$ & $\cos \theta_0 \sim U([-1,1])$ &  $\cos \theta_0 \sim U([-1,1])$ \\
Obliquity $\theta_1$ &  $\cos \theta_1 \sim U([-1,1])$ & $\cos \theta_1 \sim U([-1,1])$ \\
 Precession $\phi_0$ &   $U([0,2\pi]) $ &   $U([0,2\pi]) $\\
 Precession $\phi_1$ &   $U([0,2\pi]) $ &   $U([0,2\pi]) $ \\
\hline 
 Argument of pericenter $\omega_1$ & $U([0,2\pi])$ &$U([0,2\pi])$ \\
 Argument of pericenter $\omega_2$ & $U([0,2\pi])$ & $U([0,2\pi])$ \\
  Longitude of the ascending node $\Omega_1$ & $U([0,2\pi])$ & $U([0,2\pi])$\\
  Longitude of the ascending node $\Omega_2$ & $U([0,2\pi])$ & $U([0,2\pi])$\\
\hline 
Magnetic spin-down & Yes& Yes \\  
\hline   
\end{tabular}
}
\end{center}
\label{table_CI}
\end{table*}

\normalsize

The expressions of the vectorial secular equations are given in Appendix \ref{app} for completeness. Using these equations, 
we simulate the typical evolution of a triple-star system undergoing Lidov-Kozai migration. In Fig.~\ref{fig:evolution_systeme4856} we show the evolution of the system whose initial conditions are reported in Table~\ref{ci_systeme4856}. For the Lidov-Kozai mechanism to be initiated, we consider a large initial mutual inclination, namely $105.32^\circ$. Rapidly, we see that the inner orbit undergoes Lidov-Kozai cycles during which $e_1$ and $i$ oscillate in opposite phase and $\omega_1$ librates around $90^\circ$. Because of the strength of the dissipative tides at close distance, each time $e_1$ reaches a high value, the inner orbit loses energy and the semi-major axis $a_1$ decreases. The inner orbit finally ends in a quasi-circular orbit with orbital period of a few days only. In our example, the orbital period $P_1$ and the periods of rotation ($P_{{\rm rot},0}$ and $P_{{\rm rot},1}$) are pseudo-synchronized at $2.87$ days, while the mutual inclination $i$ stabilizes at $133.8^\circ$ at the end of the simulation. The extensive simulations carried out in this work are described in the next section.

\begin{figure*}
 	\includegraphics[width=2.\columnwidth]{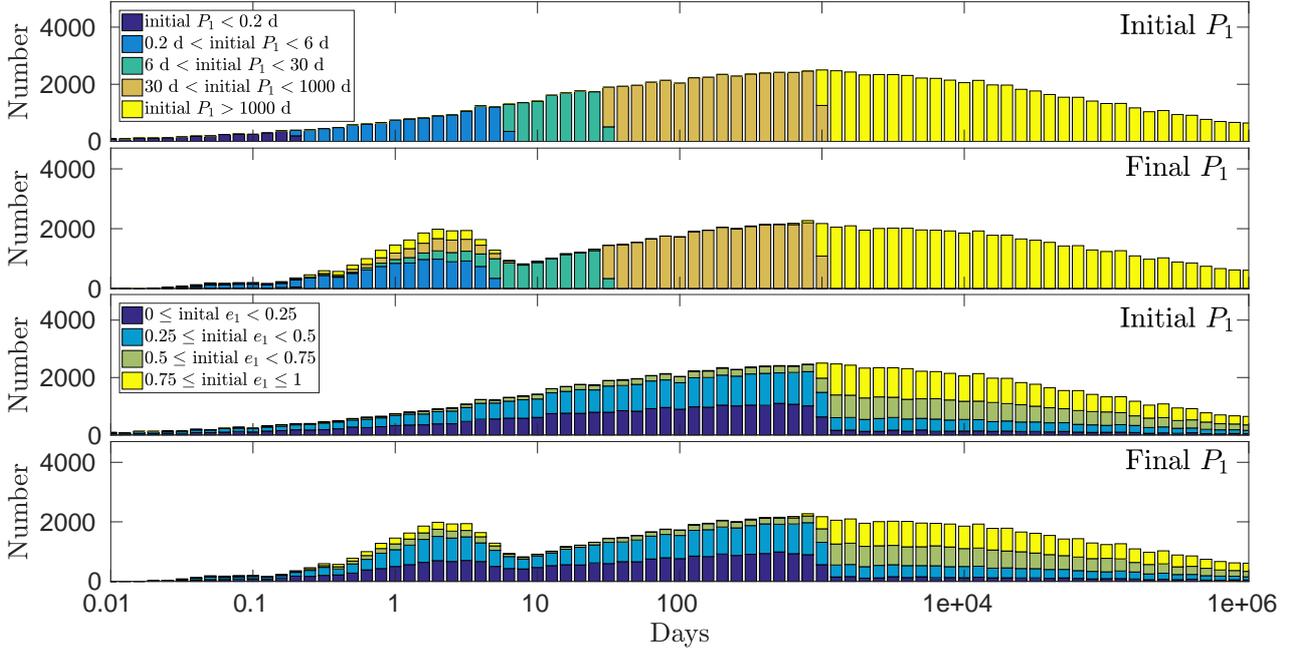}
 	  \caption{Simulations performed with the non-uniform initial distributions (second column of Table \ref{table_CI}). Top panel: Histogram of the initial orbital periods $P_1$. Second panel: Histogram of the final $P_1$, colored according to the initial $P_1$. Third panel: Histogram of the initial $P_1$, colored according to the initial eccentricity $e_1$. Bottom panel: Histogram of the final $P_1$, colored according to the initial $e_1$. The periods are displayed in logarithmic scale.}
    \label{fig:hist_periodes_periodes_e1_NU}
\end{figure*}

\begin{figure}
	\centering
	\includegraphics[width=0.9\columnwidth]{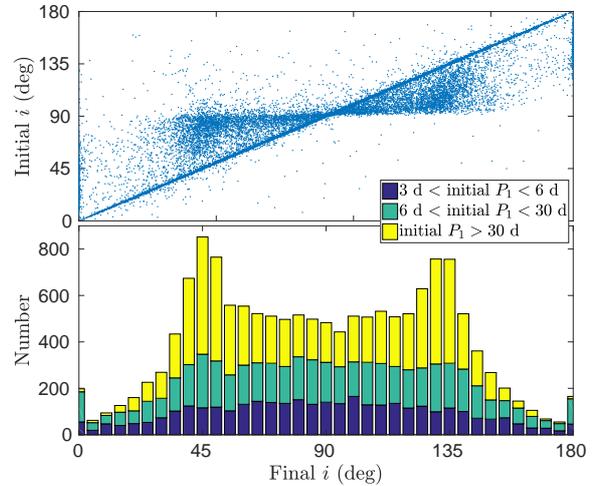}
    \caption{Top: Initial and final mutual inclinations. Bottom: Histogram of the final inclination $i$. Both plots consider only the triple-star systems with $1< {\rm final\,} P_1 < 10$ days (pile-up). Simulations performed with the non-uniform initial distributions (second column of Table \ref{table_CI}).}
    \label{fig:inclinaisons_NU}
\end{figure}

\section{Non-uniform initial distributions}\label{sec:nonuniform}




Previous studies on the evolution of triple-star systems through Lidov-Kozai migration have shown that this mechanism is a possible explanation for the three-day pile-up in the distribution of the orbital periods of multiple stars (see Section~\ref{sec:Introduction}). In this work we aim to identify the initial orbital configurations of triple-star systems leading to this kind of migration. To do so, we perform extensive numerical simulations of 100\,000 triple-star systems.
As a first step, we adopt nearly the same initial conditions as the ones from the simulations performed by \cite{Fabrycky_Tremaine_2007} (based on the observed distributions of \cite{Duquennoy_Mayor_1991}). These conditions are displayed in Table \ref{table_CI} (second column). The mass ratios follow a Gaussian normal distribution $\mathcal{N}(0.23,0.42)$. The initial distributions of the periods $P_j$ follow a log-normal distribution $\log_{10}(P_j (d)) \sim \mathcal{N}(4.8,2.3)$. Two different initial eccentricity distributions are considered depending on the initial orbital period of the star:
\begin{eqnarray}\label{ecc1}
e_1 &\sim& \text{Rayleigh}\,(\beta=0.33)  \quad {\, \rm when \,\,}  P_1< 1000  {\, \rm d}   \\ \nonumber
&=& \sqrt{-\beta^{2}\ln{(1+u(\exp{(-1/\beta^2)}-1))}}  
\end{eqnarray}
and
\begin{eqnarray}\label{ecc2}
e_1 &\sim& \text{Ambartsumian} \quad {\, \rm when \,\,}  P_1 > 1000  {\, \rm d}  \\ \nonumber 
&=&\sqrt{u}  
\end{eqnarray}
with $u\sim U([0,1])$. The mutual inclination of the third star follows an isotropic distribution: $\cos i \sim U([-1,1])$. Since we focus on hierarchical systems in this work, we only simulate systems that respect the criterion of \cite{Mardling_Aarseth_2001}:
\begin{equation}
 \frac{a_2}{a_1}> 2.8\left(1+\frac{m_2}{m_0+m_1}\right)^{2/5} (1+e_2)^{2/5} (1-e_2)^{-6/5} \left(1-\frac{0.3 i}{180^\circ}\right).
\end{equation}

\begin{figure}
	\includegraphics[width=\columnwidth]{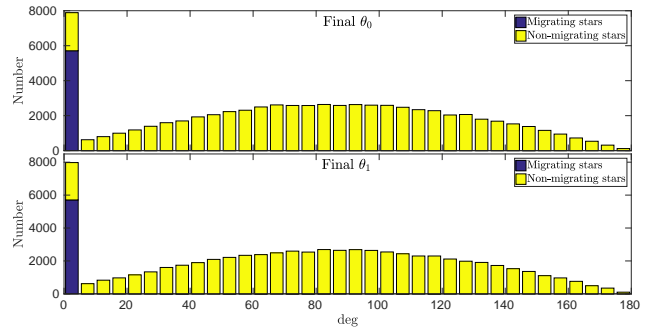}
    \caption{Final distributions of the obliquities $\theta_0$ (top) and $\theta_1$ (bottom). Simulations performed with the non-uniform initial distributions (second column of Table \ref{table_CI}).}
    \label{fig:theta0_final_NU}
\end{figure}

\begin{figure*}
	\includegraphics[width=2.0\columnwidth]{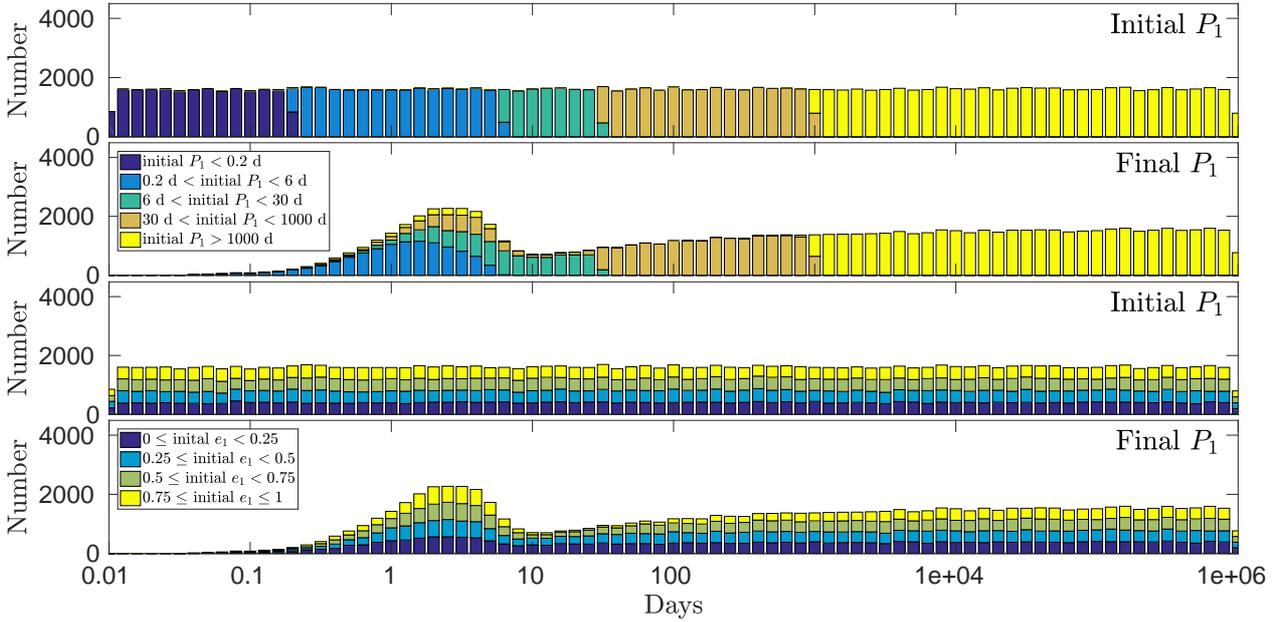}
    \caption{Same as Fig. \ref{fig:hist_periodes_periodes_e1_NU} for the uniform initial distributions (third column of Table \ref{table_CI}).}
    \label{fig:hist_periodes_periodes_U}
\end{figure*}

Unlike the work of \cite{Fabrycky_Tremaine_2007}, our study also includes octupole terms in the secular approximation, the magnetic spin-down of the two inner stars and a criterium for the disruption cases. Following \cite{Anderson_et_al_2016_mar}, the initial choice for the stellar rotation periods is $2.3$ days. 

Concerning the stopping conditions for our simulations, we implement the following ones:
\begin{itemize}
\item Disruption condition: as in \cite{Petrovich_2015_jan} and \cite{Anderson_et_al_2016_mar}, to avoid close approaches, we stop the integration when the periastron distance is smaller than the Roche limit, i.e., such that 
\begin{equation}\label{criterion}
 a_1(1-e_1) <  r_t = f_t R_1 \left( \frac{m_0}{m_0+m_1} \right)^{-1/3}.
\end{equation}
The value of $f_t$ is fixed to $2.46$, following \cite{Chandrasekhar_1997}.
\item Migration condition: to save CPU time, we no longer follow the evolution of a system when the inner orbit becomes very small and nearly circular, namely when $a_1 < 0.1$ AU and $e_1 < 10^{-4}$. 
\item CPU condition: the maximal CPU time for an integration is fixed to 6 hours. Less than $1\%$ of the simulations are concerned, and they are removed from our analysis.
\item Integration time condition: the integration time length (yr) is not a fixed value, but it follows an uniform distribution $U([0.5 \times 10^9 , 10^{10}])$, as in \cite{Petrovich_2015_jan}. However, the tidal parameter is fixed to $\Delta t_j=0.5$ sec (\citealt{Wu_Murray_2003})\footnote{To fix the tidal parameter while considering different integration time values is similar as adopting different tidal parameter values for a fixed integration time.}.
\end{itemize}

The equations (\ref{equations_vectorielles_octupole_G1})-(\ref{equations_vectorielles_octupole_Lj}) are integrated with a Bulirsch-Stoer integrator (\citealt{Stoer_Bulirsch_1980}), with a precision $10^{-6}$. The step is fixed to $100$ years initially. The total CPU time required for our simulations was $\sim 3.2\times10^4$ computational hours.


In Fig.~\ref{fig:hist_periodes_periodes_e1_NU}, we display the initial and final orbital periods of our simulations (top two panels). The general trend in the final distribution is very similar to \cite{Fabrycky_Tremaine_2007} (see their figure 5), although we took into account additional effects. The well known three-day pile-up is clearly reproduced and extends mostly from $0.5$ days to $10$ days. The color code in the top two panels groups different initial orbital period for the systems, showing that the short-period accumulation is mainly due to the migration of systems with initial orbital periods $P_1$ higher than $30$ days. In the following, we will denote {\it migrating star} a star for which initial $P_1 > 30$ days and final $P_1 < 10$ days.

\begin{figure*}
    \includegraphics[width=2.2\columnwidth]{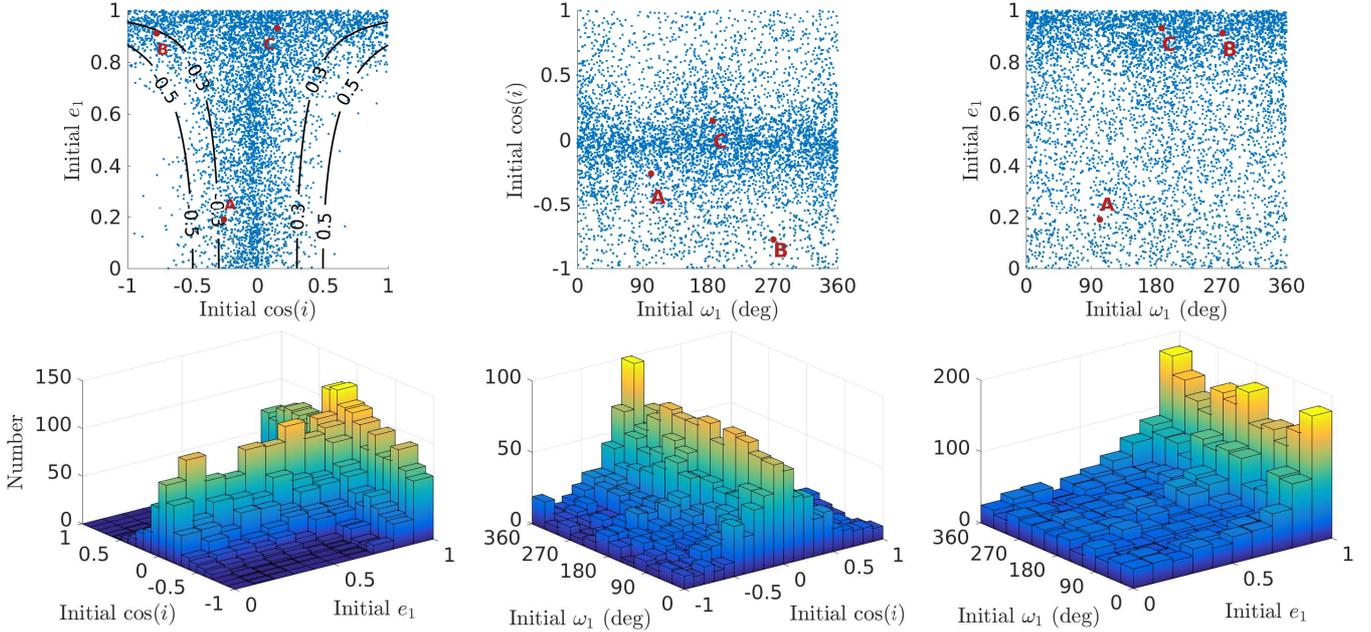}
    \caption{Initial orbital elements (top panels) and bivariate histograms (bottom panels) of the uniform initial distributions (third column of Table \ref{table_CI}) for the migrating stars only. Left panels: $(\cos i,e_1)$, middle panels: $(\omega_1,\cos i)$ and right panels: $(\omega_1,e_1)$. Black level curves of $h$-values in the top left panel are discussed in Section~\ref{sec:analytical}.}
    \label{fig:hist3_e1_i_U}
\end{figure*}

As we are interested in the initial conditions favoring the formation of the short-period pile-up, the color code of the histograms of the initial and final orbital periods $P_1$ in the bottom two panels of Fig.~\ref{fig:hist_periodes_periodes_e1_NU} refers to the initial eccentricity values $e_1$. The two different initial eccentricity distributions are clearly visible (Eqs.~(\ref{ecc1})-(\ref{ecc2})). We note that, although the eccentricities of close-in stars are low to moderate in the initial distribution ($e_1<0.5$), a significant percentage of stars with initially high eccentricity ($e_1>0.5$) are gathered in the three-day pile-up at the end of the simulations.           

Concerning the evolution of the mutual inclination of triple-star systems for which $1 <$ final $P_1 < 10$ days (short-period pile-up), 
Fig.~\ref{fig:inclinaisons_NU} shows a similar trend as Fig.~7 in \cite{Fabrycky_Tremaine_2007}. In the top panel, we show the repartition of the initial mutual inclination $i$ with respect to the final value. The majority of the systems keep their mutual inclination unchanged (first bissector). Also, a large number of systems in the short-period pile-up have initial $i \sim 90^{\circ}$ and depart from this value after the Lidov-Kozai migration. The main difference with respect to \cite{Fabrycky_Tremaine_2007} (quadrupole), also highlighted by \cite{Naoz_Fabrycky_2014} (octupole), is that the final mutual inclinations are more diversified, even leading to possible orbit flippings. In the bottom panel, the two peaks around $40^\circ$ and $140^\circ$ in the final mutual inclination distribution, previously found by \cite{Fabrycky_Tremaine_2007}, are also clearly visible. The accumulation at these two inclination values is mainly due to migrating stars (yellow).

The obliquity of the host star to the orbital plane of the inner orbit $\theta_0$ is widely discussed in the literature (e.g., \citealt{Fabrycky_Tremaine_2007, Correia_et_al_2011, Naoz_Fabrycky_2014, Anderson_et_al_2016_mar}). Although this is not the focus of the present work, we note that, similarly to \cite{Naoz_Fabrycky_2014}, for migrating star systems, the final obliquities $\theta_0$ and $\theta_1$ tend to be close to $0^\circ$ in our simulations (Fig. \ref{fig:theta0_final_NU}, blue), while the values are more diversified for non-migrating star systems (yellow), similarly to the initial distribution ($\cos \theta_j \sim U([-1,1]), \, j=0,1$).

The previous figures have shown that, although additional effects have been considered (octupole expansion, magnetic spin-down of the inner binary, disruption, different integration times), our results are consistent with the works of \cite{Fabrycky_Tremaine_2007} and \cite{Naoz_Fabrycky_2014}. In the next section, we aim to go one step further in the analysis by determining which initial values of the orbital elements initiate the migration and the subsequent formation of the three-day pile-up. In order to avoid possible biases in the selection of the initial conditions of the systems, we will adopt uniform initial distributions.


%
%



\section{Uniform initial distributions}\label{sec:uniform}


In this section, we adopt uniform initial distributions for the masses ($m_i\sim U([0.1,1.5])$), orbital periods ($\log_{10}(P_i (d)) \sim U([-2,6])$) and eccentricities ($e_i\sim U([0,1])$), $i=1,2$. The initial distributions for the other parameters are unchanged (see third column of Table \ref{table_CI}). 

In our simulations, $29\%$ of the systems are tidally disrupted, according to Eq. (\ref{criterion}). For comparison, \cite{Naoz_Fabrycky_2014} reported only $4\%$ of tidally disrupted systems. Migrating stars (initial $P_1 > 30$~days and final $P_1 < 10$ days) represent $8\%$ of the stars with initial orbital period higher than $30$~days. This percentage is the same as in  \cite{Naoz_Fabrycky_2014}.

Fig.~\ref{fig:hist_periodes_periodes_U} shows the initial and final orbital periods $P_1$ (top two panels) and the initial and final eccentricities $e_1$ (bottom two panels) for the uniform initial distributions. Regarding the eccentricities, the three-day pile-up consists of more highly eccentric orbits ($e_1>0.5$, green and yellow colors) than in the case of the non-uniform initial distributions (Fig. \ref{fig:hist_periodes_periodes_e1_NU}). The main reason is that there are more stars with orbital period between $30$ and $1000$ days (brown color) and eccentricities higher than $0.5$ (green and yellow colors) when we consider the initial uniform distributions. These stars tend to migrate, especially the highly eccentric ones.  

In the following, we aim to determine the initial values of the orbital elements leading to the migration in the three-day pile-up. In particular, we focus our efforts on the initial triplets $(e_1,\omega_1,i)$. The initial orbital elements of the migrating stars are shown in the top panels of Fig. \ref{fig:hist3_e1_i_U}, while in the bottom panels we display the 3D bivariate histograms of the initial distributions for the migrating stars, namely $(\cos i,e_1)$ (Fig. \ref{fig:hist3_e1_i_U}, left panels), $(\omega_1,\cos i)$ (Fig. \ref{fig:hist3_e1_i_U}, middle panels) and $(\omega_1,e_1)$ (Fig. \ref{fig:hist3_e1_i_U}, right panels). The yellow colors show the initial conditions for which an accumulation of migrating stars is observed in our simulations.  

An accumulation is present in the left panels of Fig.~\ref{fig:hist3_e1_i_U}, for inner orbits with initial high eccentricity ($e_1>0.8$, as previously shown in Fig.~\ref{fig:hist_periodes_periodes_U}) and initial mutual inclination around $90^\circ$. The mutual inclinations leading to migration are more restricted for moderate initial eccentricities ($i \in [60^\circ, 120^\circ]$) than for high initial eccentricities ($i \in [0^\circ, 180^\circ]$). In other words, the initial mutual inclination is close to $90^\circ$ for all initial eccentricities, except for very high eccentricity values for which all initial mutual inclination values can lead to migration. Let us note that, with the non-uniform initial distributions of the previous section, a second accumulation would be visible at moderate initial eccentricity ($e_1 \sim 0.3$) and initial mutual inclination around $90^\circ$. 

In the middle and right panels of Fig. \ref{fig:hist3_e1_i_U}, two accumulations are reported and correspond to different initial arguments of the pericenter: $\omega_1=0^\circ$ and  $\omega_1=180^\circ$. However, all initial $\omega_1$ values can induce the migration. Regardless of the initial value of $\omega_1$, migrating stars are preferably associated with an initial mutual inclination around $90^\circ$ and initial eccentricity $e_1=0.9$.

The previous analysis has shown that the initial values of the orbital parameters $e_1$, $\omega_1$ and $i$ all play an important role in the long-term dynamical evolution of a triple-star system and are crucial for the establishment of the migration. In summary, the probability to initiate the migration is higher for initial $e_1\sim 0.9$, initial $\omega_1\sim 0^\circ$ or $180^\circ$ and initial $i\sim90^\circ$. 

Three typical evolutions of migrating stars (A, B and C) are indicated in Fig. \ref{fig:hist3_e1_i_U} and are discussed in the next section.    

\section{Dynamical analysis}\label{sec:analytical}

The goal of this section is to explain the distributions observed in the previous section on the initial conditions leading to migration, by means of an analytical study of the evolutions. To understand the dynamics of the migrating stars, we resort to phase portraits of a simplified hamiltonian formulation. This helps us to follow the trajectory of the system from its initial configuration to the final one.

\subsection{Phase portraits}

\begin{figure}
	\includegraphics[width=\columnwidth]{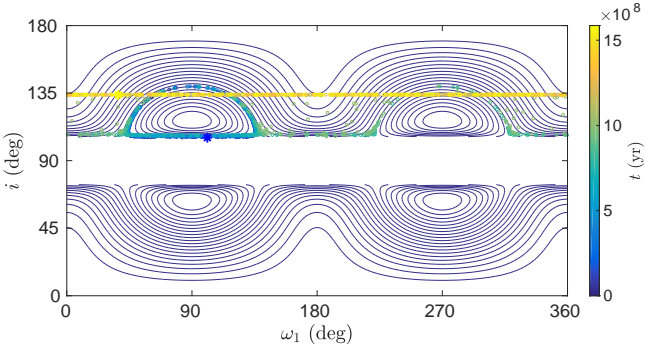}
	\caption{Phase portrait of Evolution A. Level curves of constant hamiltonian are presented in the plane ($\omega_1,i$) for the $h$-value of System A ($h=-0.26$). The time evolution is given by the colorbar. The blue star denotes the initial parameters of Evolution A, while the yellow one its final parameters.}
	\label{fig:plandephase_systeme4856}
\end{figure}

As a simplified model, we only consider the dominating conservative terms of the secular hamiltonian (after averaging over the mean anomalies), namely the quadrupolar interaction with general relativity corrections (e.g., \citealt{Correia_et_al_2013}):
\begin{equation}
 {\mathcal H} =  \frac{\varepsilon_0}{\sqrt{1-e_1^2}} + \left(2+3 e_1^2\right)\left(2- 3 \sin^2 i \right)  
  + 15 e_1^2 \sin^2 i \cos 2\omega_1 \ ,
\end{equation}
where 
\begin{equation}
\varepsilon_0= 3 \frac{m_0+m_1}{m_2} \left(\frac{n_1 a_1}{c}\right)^2 \left(\frac{a_2}{a_1}\right)^3 (1-e_2^2)^{3/2} 
\end{equation}
corresponds to the general relativity contribution.
Since the argument of pericenter of $m_2$ does not appear in the previous hamiltonian, the eccentricity of the outer orbit is constant. For hierarchical systems, the angular momentum of the outer orbit is much greater than that of the inner binary, so the adimensional angular momentum of the inner orbit, $h=\sqrt{1-e_1^2}\cos{i}$, is nearly constant (e.g., \citealt{Correia_et_al_2011}), which is usually known by the ``Kozai constant''. Therefore, in the quadrupolar approach, the hamiltonian has only one degree of freedom ($\omega_1, e_1$): 
\begin{equation} \label{hamomg}
 {\mathcal H} =  \frac{\varepsilon_0}{x} - \left(5-3 x^2\right)\left(1 - 3\frac{h^2}{x^2} \right)
  + 15 (1-x^2) \left(1- \frac{h^2}{x^2} \right) \cos 2\omega_1 \ ,
\end{equation}
with $x = \sqrt{1-e_1^2}$.
For a given value of $h$, the phase portrait of the dynamics will then consist of the level curves of the hamiltonian in the plane ($\omega_1, e_1$) or ($\omega_1,i$) (e.g., \citealt{Giuppone_et_al_2012}).

%
Only stellar systems that undergo close encounters can experience strong tidal dissipation.
As a result, the inner orbit maximal eccentricity is a critical parameter for Lidov-Kozai migration.
In a Lidov-Kozai cycle, the maximal eccentricity is always reached for $\omega_1 = 90^\circ$ or~$270^\circ$ \citep{Lidov_1962, Kozai_1962}, that is, $\cos 2 \omega_1 = -1$. For a given set of initial conditions $(\omega_{10}, e_{10}, h)$, we can then predict the maximal eccentricity using the hamiltonian (\ref{hamomg}) (when $\varepsilon_0=0$, i.e. no general relativity), by solving the following polynomial equation for $x$:

\begin{equation}\label{eqmax}
- \left(5-3 x^2\right)\left(x^2 - 3 h^2  \right)
  - 15 (1-x^2) \left(x^2- h^2 \right) =  {\mathcal H}_0  \ ,
\end{equation}
with
\begin{equation} 
 {\mathcal H}_0 =  - \left(5-3 x_0^2\right)\left(x_0^2 - 3 h^2  \right)
  + 15 (1-x_0^2) \left(x_0^2- h^2 \right) \cos 2\omega_{10} \ ,
\end{equation}
and $x_0 = \sqrt{1-e_{10}^2}$. Maximal eccentricity values reached during the Lidov-Kozai migration will be further discussed in Section~\ref{sec:h}. 

\subsection{Typical evolutions}

By a thorough analysis of the long-term evolutions related to the accumulations observed in Fig. \ref{fig:hist3_e1_i_U}, we have identified three typical dynamical behaviors among the migrating stars. These three evolutions are denoted A, B, C on Fig. \ref{fig:hist3_e1_i_U} and are detailed in the following paragraphs.   

\begin{table}
\caption{Initial conditions of Evolution B.}
\begin{center}
{\scriptsize
\begin{tabular}{|r|r|r|r|r|r|r|}
\hline
 Variables & \multicolumn{2}{r|}{$m_0$} & \multicolumn{2}{r|}{$m_1$} & \multicolumn{2}{r|}{$m_2$} \\
 \hline
 Mass $m_j$ ($m_\odot$) & \multicolumn{2}{r|}{$1$} &  \multicolumn{2}{r|}{$1.32$} &  \multicolumn{2}{r|}{$1.46$} \\
 Radius $R_j$ (m)&  \multicolumn{2}{r|}{$696.34 \times 10^6$} &  \multicolumn{2}{r|}{$871.17 \times 10^6$} &  \multicolumn{2}{r|}{-}\\
 Rotation period $P_{{\rm rot},j}$ (d)&  \multicolumn{2}{r|}{$2.3$} &  \multicolumn{2}{r|}{$2.3$} &  \multicolumn{2}{r|}{-} \\
 Obliquity $\theta_j$ (deg)&  \multicolumn{2}{r|}{$26.59$} &  \multicolumn{2}{r|}{$69.91$} &  \multicolumn{2}{r|}{-} \\
 Precession $\phi_j$ (deg)&  \multicolumn{2}{r|}{$241.25$} &  \multicolumn{2}{r|}{$42.50$} &  \multicolumn{2}{r|}{-} \\
 $C_j$ &  \multicolumn{2}{r|}{$0.08 m_0 R_0^2$} &  \multicolumn{2}{r|}{$0.08 m_1 R_1^2$} &  \multicolumn{2}{r|}{-} \\
 Love number $k_{2_j}$ &  \multicolumn{2}{r|}{$0.028$} &  \multicolumn{2}{r|}{$0.028$} &  \multicolumn{2}{r|}{-} \\
 $\Delta t_j$ (s)&  \multicolumn{2}{r|}{$0.5$} &  \multicolumn{2}{r|}{$0.5$} &  \multicolumn{2}{r|}{-} \\
\hline
 Semi-major axis $a_j$ (AU)& \multicolumn{2}{r|}{} & \multicolumn{2}{r|}{$0.63$} & \multicolumn{2}{r|}{$78.46$} \\
 Eccentricity $e_j$ & \multicolumn{2}{r|}{} & \multicolumn{2}{r|}{$0.91$} & \multicolumn{2}{r|}{$0.40$} \\
 Argument of pericenter $\omega_j$ (deg)& \multicolumn{2}{r|}{} & \multicolumn{2}{r|}{$271.58$} & \multicolumn{2}{r|}{$24.06$}  \\
 Longitude of the node $\Omega_j$ (deg)& \multicolumn{2}{r|}{} & \multicolumn{2}{r|}{$67.44$} & \multicolumn{2}{r|}{ $141.03$}  \\
\hline
 Mutual inclination $i$ (deg)&\multicolumn{3}{c|}{} &  \multicolumn{3}{c|}{$\qquad 141.73$}    \\
\hline
\end{tabular}
}
\end{center}
\label{tab:ci_systeme70426}
\end{table}

\paragraph*{Evolution A.}
The dynamical evolution of System A was previously shown in Fig. \ref{fig:evolution_systeme4856} and consists in a typical evolution of a triple-star system undergoing Lidov-Kozai migration. The initial conditions of Evolution~A are $e_1=0.19$, $i=105.32^\circ$ and $\omega_1=101.06^\circ$ (see Table \ref{ci_systeme4856} for the full list of the initial parameters). As observed in Fig.~\ref{fig:hist3_e1_i_U}, these initial conditions are suitable for migration, although the eccentricity of Evolution A is well below the accumulation at very high eccentricities observed in the uniform distribution simulations. During the Lidov-Kozai cycles, $e_1$ and $i$ oscillate in opposite phase and $\omega_1$ librates around $90^\circ$, as previously observed in Fig. \ref{fig:evolution_systeme4856}.

Another way to represent the evolution of System A is shown in Fig.~\ref{fig:plandephase_systeme4856}, where the level curves of constant energy in the plane ($\omega_1,i$) are presented for the value of $h$ corresponding to the initial conditions of the system ($h=-0.26$). This phase portrait consists of the level curves compatible with the given value of $h$, therefore for Evolution A no level curves exist between $75^\circ$ and $105^\circ$. The phase portrait clearly shows the Lidov-Kozai equilibria at $\omega_1=90^\circ$ and $\omega_1=270^\circ$. 
The color bar gives the time of the system evolution. The initial configuration of Evolution A is marked with a blue symbol, while the final configuration with a yellow one. At the beginning of the integration, the system evolves at proximity of a Lidov-Kozai equilibrium with its argument of pericenter librating around $90^\circ$ (light blue dots). At $7.5\times 10^8$ yr (green dots), due to the action of dissipative tides, Evolution A leaves out of the influence of the Lidov-Kozai equilibria ($\omega_1$ starts to circulate) and continually switches from hamiltonian curve (no energy conservation). The system is finally captured at a mutual inclination of $133.8^\circ$ (yellow dots).               

\begin{figure}
    \includegraphics[width=\columnwidth]{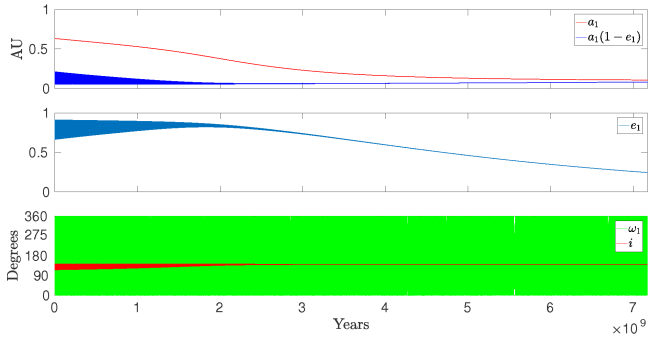}
    \caption{Evolution B. Initial parameters are given in Table \ref{tab:ci_systeme70426}.} 
    \label{fig:evolution_systeme70426}
\end{figure}

\begin{figure}
    \includegraphics[width=\columnwidth]{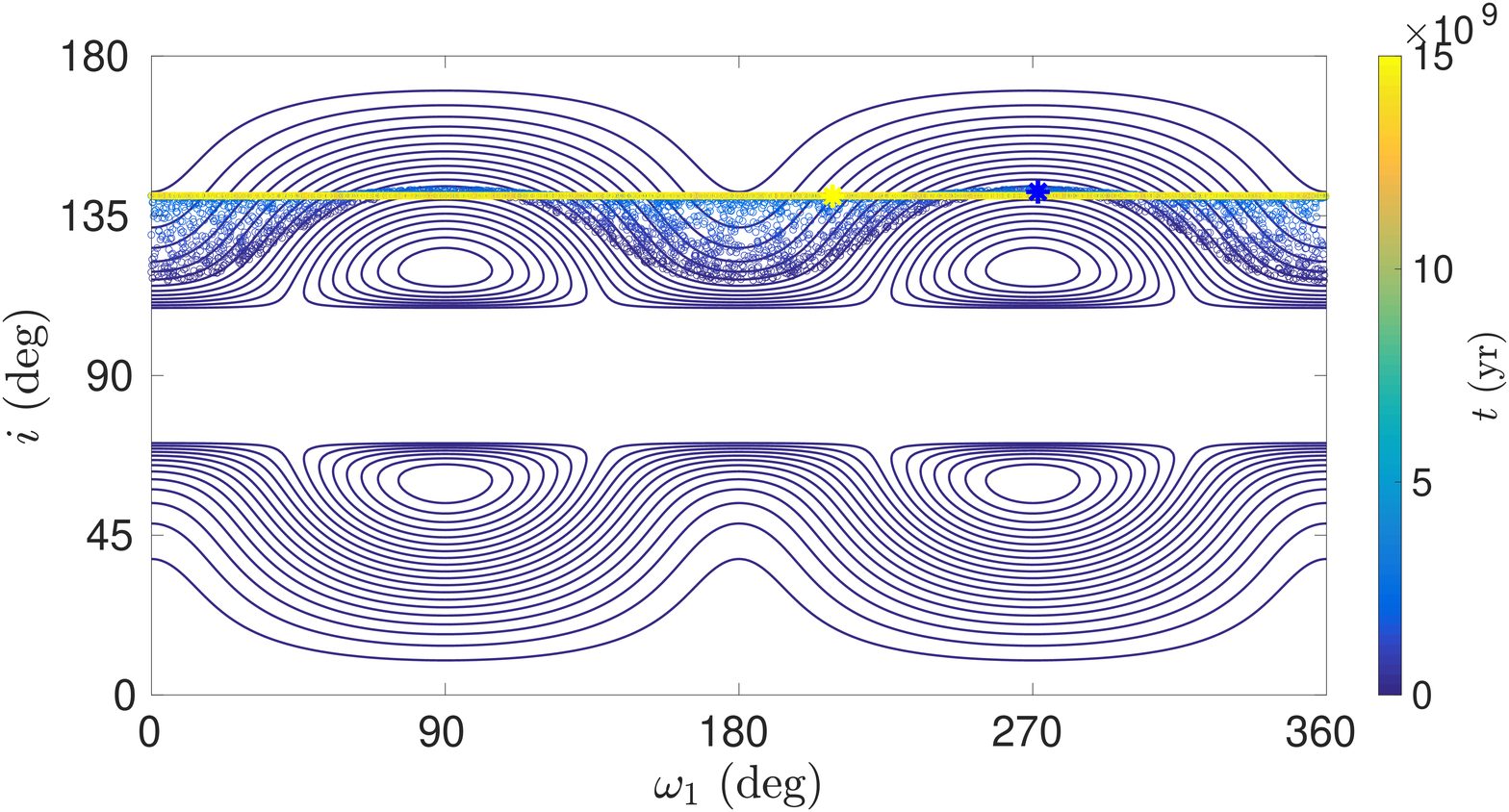}
    \caption{Phase portrait of Evolution B, as in Fig. \ref{fig:plandephase_systeme4856}.}
    \label{fig:plandephase_systeme70426}
\end{figure}

\begin{table}
\caption{Initial conditions of Evolution C.}
\begin{center}
{\scriptsize
\begin{tabular}{|r|r|r|r|r|r|r|}
\hline
 Variables & \multicolumn{2}{r|}{$m_0$} & \multicolumn{2}{r|}{$m_1$} & \multicolumn{2}{r|}{$m_2$} \\
 \hline
 Mass $m_j$ ($m_\odot$)& \multicolumn{2}{r|}{$1$} &  \multicolumn{2}{r|}{$0.41$} &  \multicolumn{2}{r|}{$0.7$} \\
 Radius $R_j$(m) &  \multicolumn{2}{r|}{$696.34 \times 10^6$} &  \multicolumn{2}{r|}{$342.04 \times 10^6$} &  \multicolumn{2}{r|}{-}\\
 Rotation period $P_{{\rm rot},j}$ (days) &  \multicolumn{2}{r|}{$2.3$} &  \multicolumn{2}{r|}{$2.3$} &  \multicolumn{2}{r|}{-} \\
 Obliquity $\theta_j$ (deg)&  \multicolumn{2}{r|}{$115.9$} &  \multicolumn{2}{r|}{$97.27$} &  \multicolumn{2}{r|}{-} \\
 Precession $\phi_j$ (deg)&  \multicolumn{2}{r|}{$156.52$} &  \multicolumn{2}{r|}{$203.88$} &  \multicolumn{2}{r|}{-} \\
 $C_j$ &  \multicolumn{2}{r|}{$0.08 m_0 R_0^2$} &  \multicolumn{2}{r|}{$0.08 m_1 R_1^2$} &  \multicolumn{2}{r|}{-} \\
 Love number $k_{2_j}$ &  \multicolumn{2}{r|}{$0.028$} &  \multicolumn{2}{r|}{$0.028$} &  \multicolumn{2}{r|}{-} \\
 $\Delta t_j$ (s)&  \multicolumn{2}{r|}{$0.5$} &  \multicolumn{2}{r|}{$0.5$} &  \multicolumn{2}{r|}{-} \\
\hline
 Semi-major axis $a_j$ (AU)& \multicolumn{2}{r|}{} & \multicolumn{2}{r|}{$3.17 $} & \multicolumn{2}{r|}{$164.12$} \\
 Eccentricity $e_j$ & \multicolumn{2}{r|}{} & \multicolumn{2}{r|}{$0.93$} & \multicolumn{2}{r|}{$0.4$} \\
 Argument of pericenter $\omega_j$ (deg)& \multicolumn{2}{r|}{} & \multicolumn{2}{r|}{$187.40$} & \multicolumn{2}{r|}{$88.27$}  \\
 Longitude of the node $\Omega_j$ (deg) & \multicolumn{2}{r|}{} & \multicolumn{2}{r|}{$104.08$} & \multicolumn{2}{r|}{ $208.67$}  \\
\hline
Mutual inclination $i$ (deg)&\multicolumn{3}{c|}{} &  \multicolumn{3}{c|}{$\qquad 81.24$}    \\ 
\hline
\end{tabular}
}
\end{center}
\label{tab:ci_systeme51993}
\end{table}

\begin{figure}
    \includegraphics[width=\columnwidth]{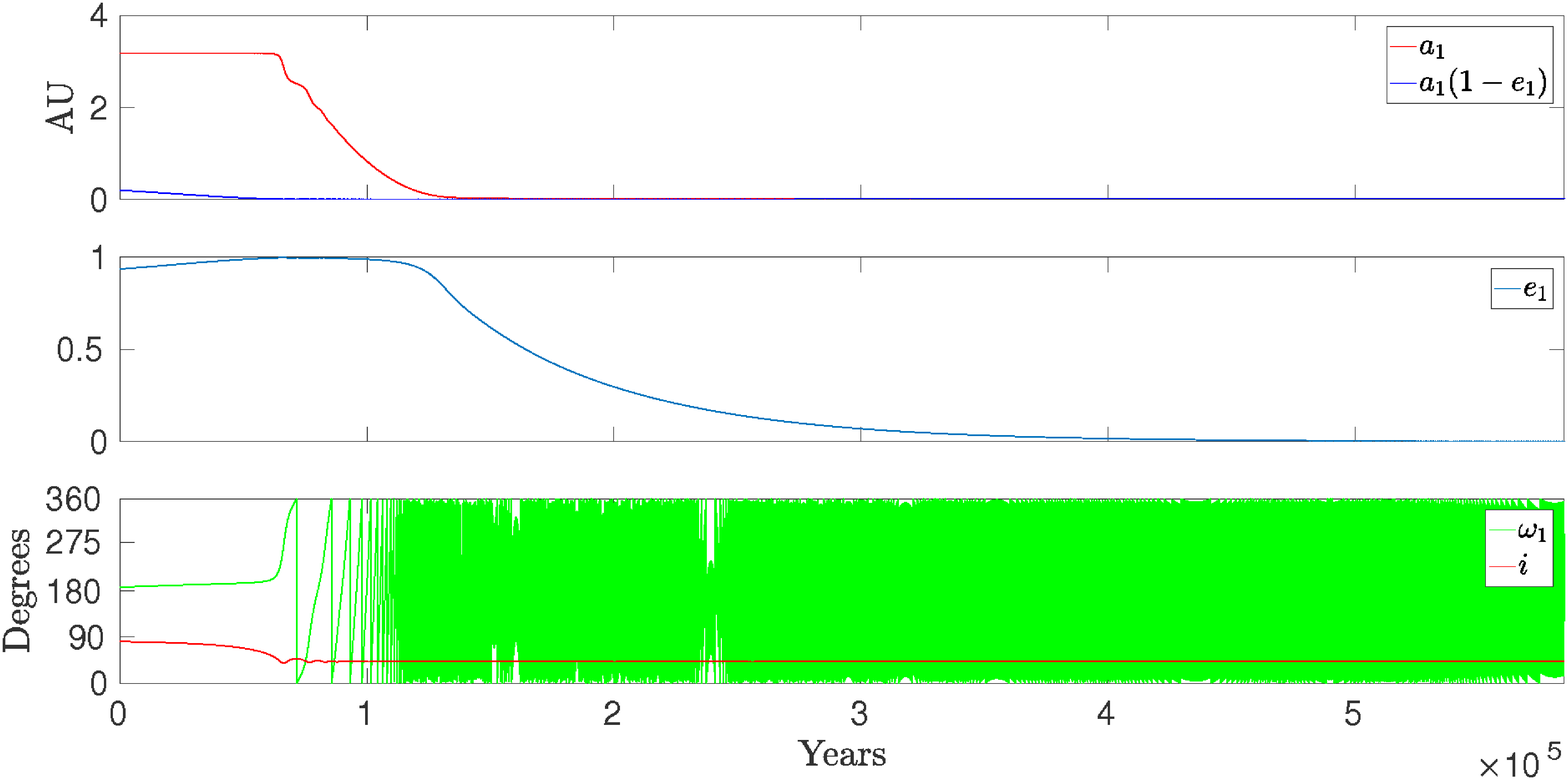}	
    \caption{Evolution C. Initial parameters are given in Table \ref{tab:ci_systeme51993}.}
    \label{fig:evolution_systeme51993_e1init}
\end{figure}

\begin{figure}
    \includegraphics[width=\columnwidth]{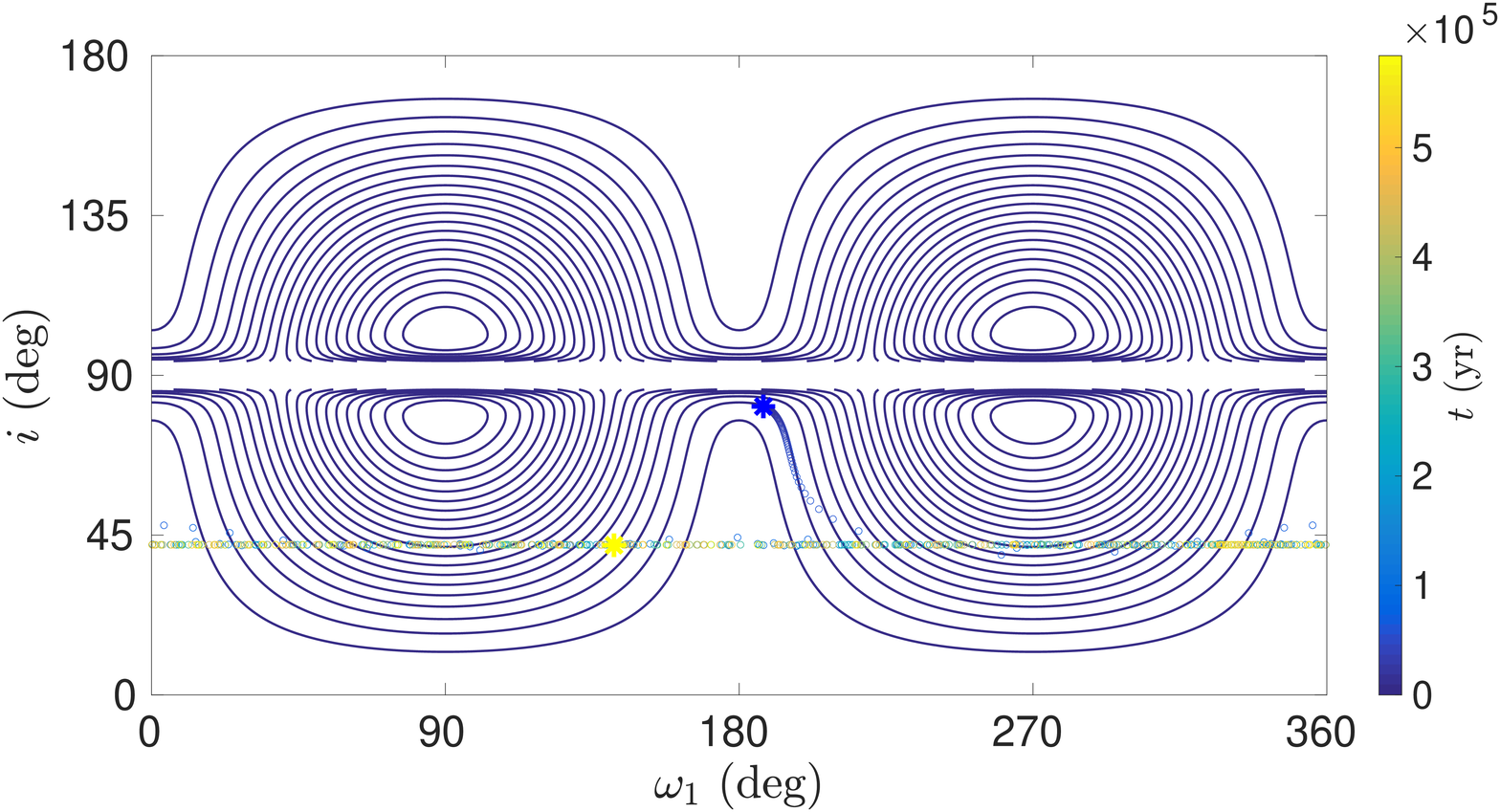}
    \caption{Phase portrait of Evolution C, as in Fig. \ref{fig:plandephase_systeme4856}.}
    \label{fig:plandephase_systeme51993_e1init}
\end{figure}

\begin{figure}
	\includegraphics[width=\columnwidth]{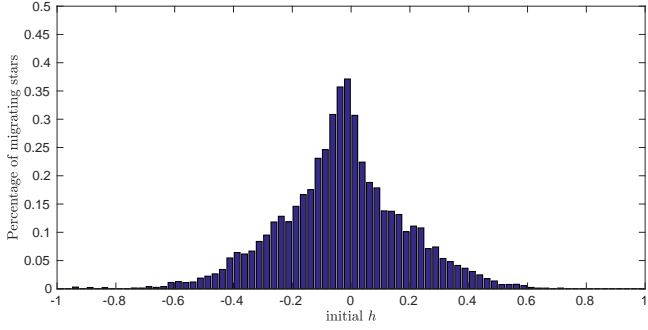}
    \caption{Histogram of the percentage of migrating stars as a function of the initial $h$-value. The peak is associated to the maximal value $h=-0.014$. Simulations performed with the uniform initial distributions.}
    \label{fig:moment_cinetique_U}
\end{figure}

\paragraph*{Evolution B.}

A second typical dynamical behavior is shown in Fig. \ref{fig:evolution_systeme70426}. Table~\ref{tab:ci_systeme70426} gathers the initial conditions of Evolution B. In particular, we see that the initial inner orbit is highly eccentric, while the mutual inclination is further away from the accumulation at $i=90^\circ$ observed in Fig. \ref{fig:hist3_e1_i_U} ($e=0.91$, $i=141^\circ$ and $\omega_1=271^\circ$). Unlike Evolution A, the inner argument of the pericenter circulates, thus the system is not inside the Lidov-Kozai resonance. By analyzing the phase portrait for $h=-0.33$ in Fig.~\ref{fig:plandephase_systeme70426}, we see that Evolution B is indeed not close to the Lidov-Kozai equilibria. However, the evolution of System B is still influenced by these equilibria, since it follows the level curves of constant hamiltonian that are distorted by the presence of the equilibria (blue dots). It explains the limited variations in eccentricity and inclination observed at the beginning of the evolution (see Fig. \ref{fig:evolution_systeme70426}). Since the initial eccentricity is very high, the strength of the dissipative tides at close distance finally initiates the migration of the system to a quasi-circular orbit with orbital period of a few days only (yellow dots).

\paragraph*{Evolution C.}

The last system gathers together a very high initial eccentricity ($e_1=0.93$), a mutual inclination close to $90^\circ$ ($i=81.24^\circ$) and an argument of the periastron close to $180^\circ$ ($\omega_1=187.40^\circ$) (see Table \ref{tab:ci_systeme51993} for the full parameters). It belongs to the accumulation of migrating star systems observed in the previous section (yellow peak in Fig. \ref{fig:hist3_e1_i_U}). The evolution of System C is shown in Fig. \ref{fig:evolution_systeme51993_e1init}. The circularisation of the inner orbit occurs on a very short timescale (less than $5\times10^5$ yr). We display in Fig. \ref{fig:plandephase_systeme51993_e1init} the phase portrait for $h=0.056$. We observe that the system is initially far from the Lidov-Kozai equilibria and barely has the time to follow the level curve of constant hamitonian before the strong tidal dissipation comes into play.    

In conclusion, the three typical evolutions presented here illustrate the influence of the Lidov-Kozai mechanism on the migration process. In particular, the proximity to the Lidov-Kozai equilibria shapes the amplitude of the eccentricity/inclination variations of the inner orbit. We have shown that for migration, very high orbital eccentricities are needed. Therefore, either the inner orbit is initially very eccentric (like Evolutions B and C), or its eccentricity has to be pumped to high values via the Lidov-Kozai cycles (like Evolution A). In the latter case, the initial mutual inclinations need to be in the range $[\arccos{(-\sqrt{3/5})},\arccos{(\sqrt{3/5})}]$ \citep{Lidov_1962, Kozai_1962}. These two possibilities explain the T-shaped structure observed in the top left panel of Fig.~\ref{fig:hist3_e1_i_U}. Let us note that the final mutual inclinations observed in the three evolutions belong to the two peaks around $40^\circ$ and $140^\circ$ in the final mutual inclination distribution observed in Fig.~\ref{fig:inclinaisons_NU}.

\begin{figure}
	\includegraphics[width=1.\columnwidth]{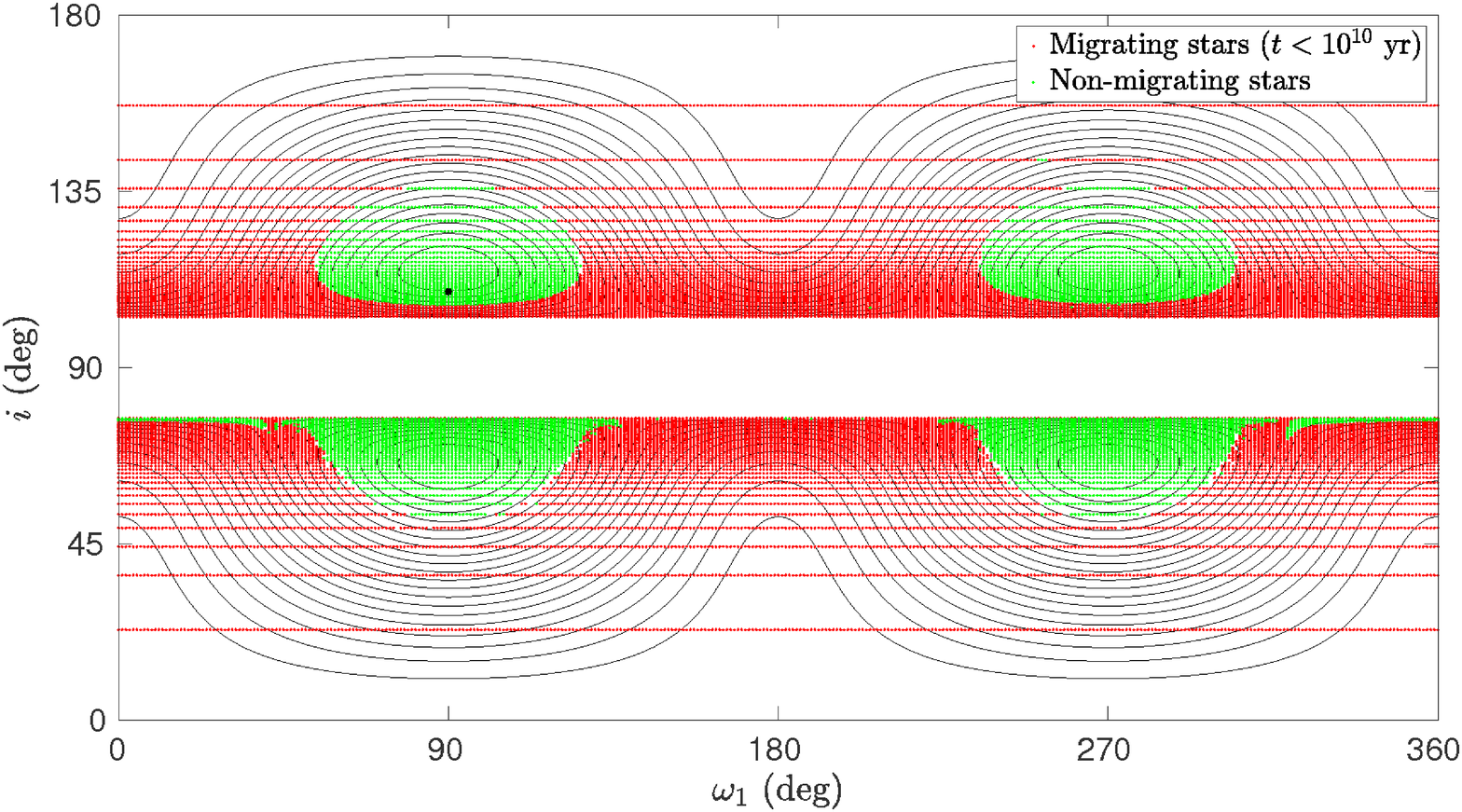}
%

\includegraphics[width=1.\columnwidth]{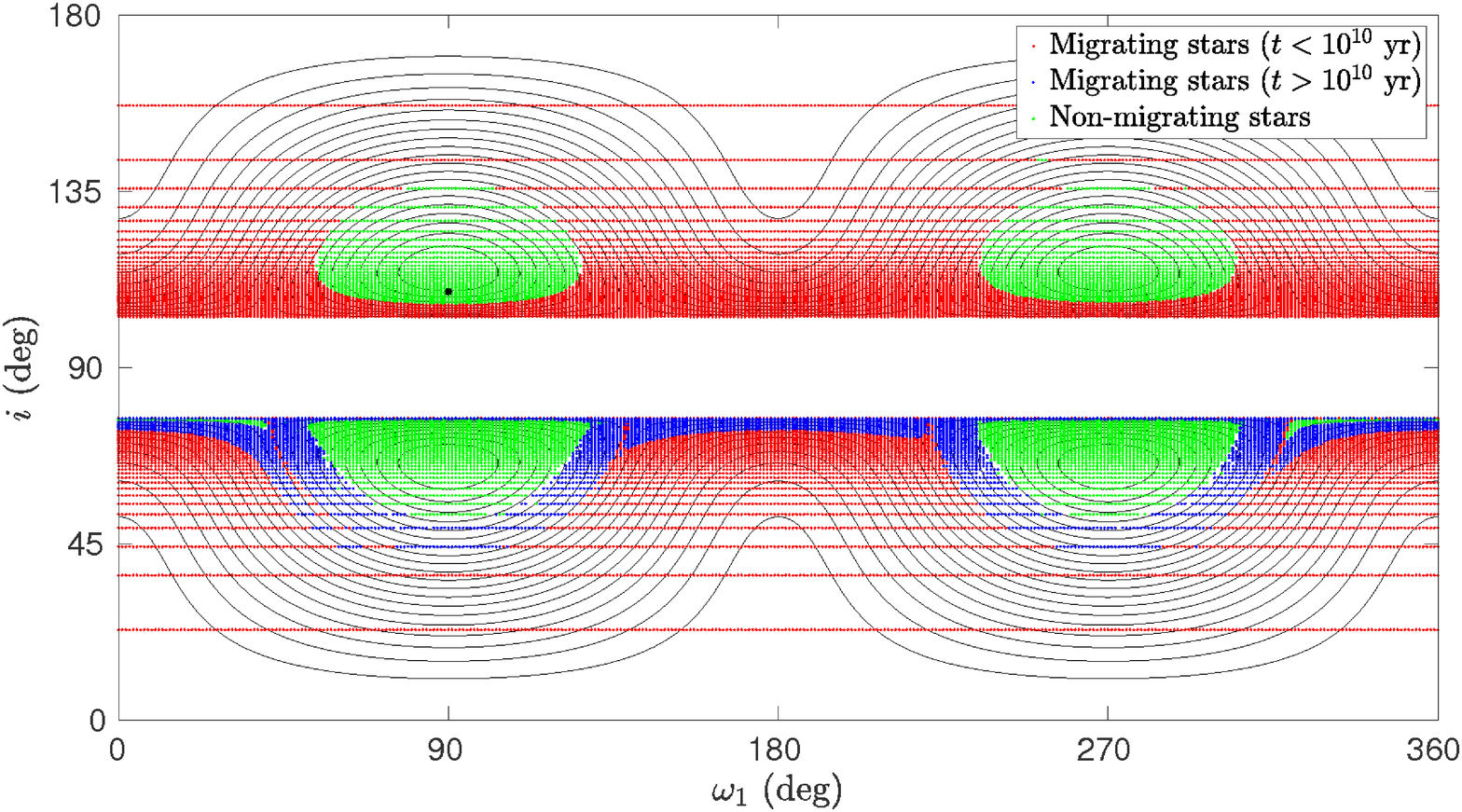}

\includegraphics[width=1.\columnwidth]{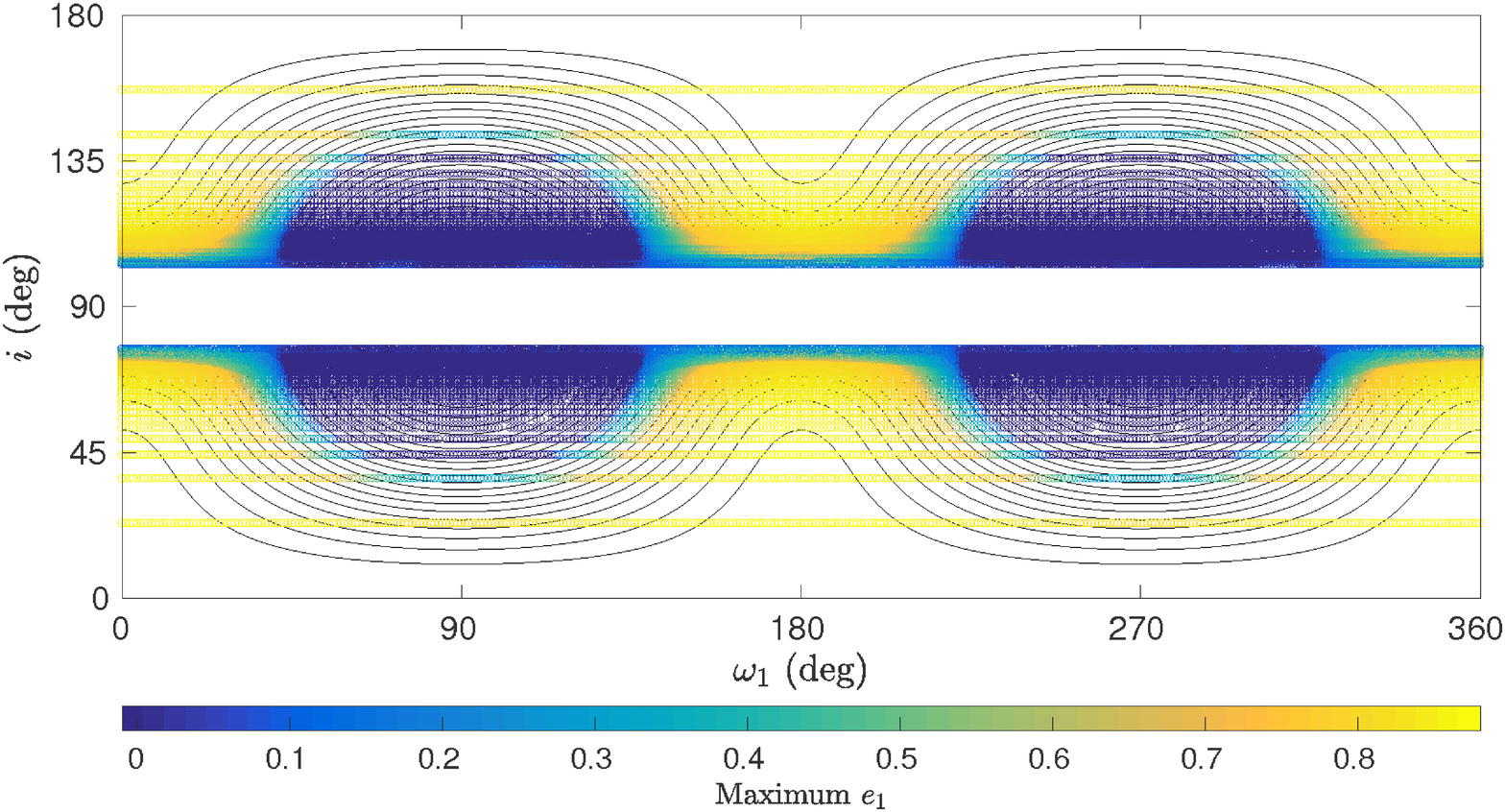}
    \caption{Top and middle panels: Initial conditions of $70\,000$ systems reported on their phase portrait ($|h|=0.224$). The integration time of the simulations is $5 \times 10^{9}$ yr in the top panel and $5 \times 10^{10}$ yr in the middle panel. Red dots stand for systems that migrate in less than $ 10^{10}$ yr, blue dots in less than $5 \times 10^{10}$ yr, and green dots for non-migrating stars. The black dot identifies the non-migrating system whose evolution is shown in Fig.~\ref{fig:evolutionD}. Bottom panel: Maximal eccentricity value of the inner orbit, as given by solving Eq.~(\ref{eqmax}).} 
    \label{fig:h5classeF}
\end{figure}

\section{Initial conditions for migrating stars}\label{sec:h}

In the previous section, three evolutions of migrating stars have been shown. Their initial $h$-values are all small ($|h| \leq \sim 0.3$). To figure out the range of initial $h$-values promoting the migration, we show in Fig.~\ref{fig:moment_cinetique_U} the histogram of the percentage of migrating stars as a function of the initial $h$-value for the initial uniform distributions. 
We observe that the $h$-values (in absolute value) of migrating stars are all gathered well below $0.5$. This is a necessary but not sufficient condition for the migration, since the initial $h$-value of non-migrating stars can also be small. However, the smaller the initial $h$-value, the higher the percentage of migrating stars. To emphasize this point, the level curves for $h=0.3$ and $h=0.5$ are plotted in the top left panel of Fig. \ref{fig:hist3_e1_i_U}. They perfectly match the T-shaped structure observed in Fig.~\ref{fig:hist3_e1_i_U} for the migrating stars, showing the importance of the initial $h$-value on the migration process.  

Furthermore, we also note that the peak of migrating stars in the histogram of Fig.~\ref{fig:moment_cinetique_U} is not centered around $h=0$, but around $h=-0.014$. Thus, the migration process tends to be more efficient for initial retrograde orbits ($h<0$).

\begin{figure}
	\includegraphics[width=\columnwidth]{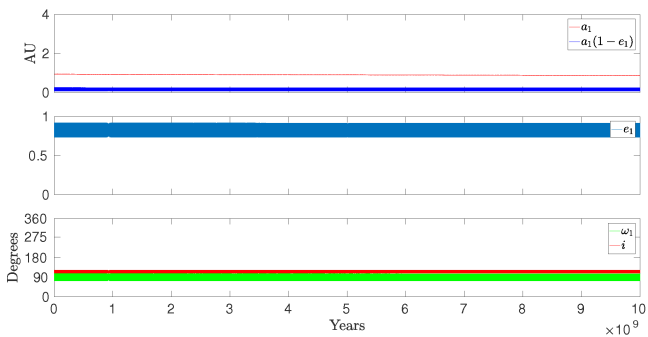}
    \caption{Typical evolution of a non-migrating system located close to a Lidov-Kozai equilibrium. The initial parameters of the system are $e_1=0.74$, $i=109.40^\circ$ and $\omega_1=90^\circ$, all the others being the same as Evolution A (Table~\ref{ci_systeme4856}).}
    \label{fig:evolutionD}
\end{figure}
 
To identify more precisely the initial orbital parameters leading to migration, a new simulation of $70 \,000$ systems is performed for $5$~Gyr. The initial parameters of the systems are similar to the ones of Evolution A, but a large variety of initial $e_1$ and $\omega_1$ values are considered (from $0$ to $0.97$ every $0.01$ and from $0^\circ$ to $360^\circ$ every $1^\circ$, respectively). For all systems, we fix initial $|h|=0.224$ (value compatible with migration, see Fig.~\ref{fig:moment_cinetique_U}). Thus, the inclination values are varied accordingly to keep $h$ fixed. 

We display in Fig.~\ref{fig:h5classeF} the phase portrait of the level curves of constant hamiltonian for $|h|=0.224$. The initial conditions of our simulations leading to migration are plotted in red on the phase portrait (top panel). We see that a large number of initial conditions lead to migration. Non-migrating stars include configurations very close to the centers of the Lidov-Kozai equilibria (green dots). An example of such an evolution, identified by the black dot in Fig.~\ref{fig:h5classeF}, is given in Fig.~\ref{fig:evolutionD}. The initial parameters of the system are  $e_1=0.74$, $i=109.40^\circ$ and $\omega_1=90^\circ$. The variation in the inner eccentricity is limited and the tidal effects do not come into play. This explains that accumulations of migrating stars were previously observed in Fig.~\ref{fig:hist3_e1_i_U} around initial $\omega_1=0^\circ$ and $180^\circ$, and not around $90^\circ$ and $270^\circ$. 

Top panel of Fig.~\ref{fig:h5classeF} also shows an asymmetry between the prograde (initial $i<90^\circ$) and retrograde (initial $i>90^\circ$) migrating stars. This assymetry has previously been observed in Fig.~\ref{fig:moment_cinetique_U}, where the peak of migration stars is observed for retrograde systems (initial $h<0$). To understand this feature, we have rerun the simulations of the non-migrating systems for a longer integration time of $50$ Gyr and our results are shown in the middle panel of Fig.~\ref{fig:h5classeF}. We display with blue dots the systems that migrate with this longer timescale. A symmetry between the initial prograde and retrograde systems is now observed. As a result, we conclude that the migration timescale depends on the initial mutual inclination of the triple-star system.

Finally, in the bottom panel of Fig.~\ref{fig:h5classeF}, we show, for the $70 \,000$ systems, the maximal eccentricity values of the inner orbit during the Lidov-Kozai cycles, as predicted from our simplified hamiltonian formulation~(\ref{hamomg}). As expected, the migrating stars are associated with the higher maximal eccentricities, which induce close encounters and thus strong tidal dissipation during the system evolution.

\section{Conclusions}\label{sec:ccl}

In this work, we have revisited the Lidov-Kozai migration with the aim of identifying the initial configurations of triple-star systems that promote the formation of the short-period pile-up present in the observations of multiple systems. Using the octupole order approximation with general relativity corrections, stellar oblateness, tides and magnetic spin-down braking, we have performed a statistical analysis of one hundred thousand simulations. This analysis has emphasized the importance of the choice of the initial distributions of the stellar orbital parameters on the final results. By adopting initial uniform distributions, we have shown that while the mutual inclination of the triple-star system is a critical parameter, the eccentricity as well as the argument of the pericenter of the inner binary also play an important role in the establishment of the migration. In particular, unbiaised initial conditions leading to migration were discussed, showing that the probability to initiate the migration is higher for initial $e_1\sim0.9$, initial $\omega_1\sim 0^\circ$ or $180^\circ$ and initial $i\sim90^\circ$. While larger ranges of these initial parameters support the migration, the only necessary condition is $ i \in [\arccos{(-\sqrt{3/5})},\arccos{(\sqrt{3/5})}]$ \citep{Lidov_1962, Kozai_1962}.

To understand the dynamics of the migrating stars, we have taken a close look at three different typical evolutions and computed phase portraits of the quadrupolar hamiltonian with general relativity corrections, where the trajectory of the systems could be followed from their initial configuration to the final one. We show that the proximity to the Lidov-Kozai equilibria shapes the amplitude of the eccentricity/inclination variations of the inner orbit and is a crucial factor for the migration. Since very high eccentricities are needed for the migration, either the inner binary is initially formed on very eccentric orbit (like Evolutions B and C), or its eccentricity has to be pumped to high values via the Lidov-Kozai cycles (like Evolution A). However, when the system is very close to the centers of the Lidov-Kozai equilibria, its variations in eccentricity are too limited for the tidal effects to come into play. Finally, the importance of the low initial $h$-value (i.e., initial $|h|<0.5$) in the dynamical evolution of migrating stars has also been highlighted. 

Although the Lidov-Kozai migration is a robust mechanism to produce the pile-up around three-day periods, it should be noted that the initial conditions for the establishment of this mechanism are demanding. It remains to be shown that such conditions can be encountered during the formation of triple-star systems.

\section*{Acknowledgements}
The authors would like to thank E.~Bolmont for useful discussions. Computational resources have been provided by the Consortium des \'{E}quipements de Calcul Intensif (C\'{E}CI), funded by the Fonds de la Recherche Scientifique de Belgique (F.R.S.-FNRS) under Grant No. 2.5020.11. A.C. acknowledges support from CIDMA strategic project (UID/MAT/04106/2013), ENGAGE SKA (POCI-01-0145-FEDER-022217), and PHOBOS (POCI-01-0145-FEDER-029932), funded by COMPETE 2020 and FCT, Portugal.



\bibliographystyle{mnras}
\bibliography{mabibliographie}



\appendix

\section{Equations of motion}\label{app}

The orbital and spin evolutions are followed using the vectorial secular equations of \cite{Correia_et_al_2016}, averaged over the mean anomalies and which include general relativity corrections, conservative and dissipative tides, stellar oblateness and magnetic spin-down braking. Their expressions are

\footnotesize
\begin{equation}\label{equations_vectorielles_octupole_G1}
\begin{aligned}
\dot{\textbf{G}}_1=&\underbrace{-\gamma (1-e_1^2) \cos i \, \hat{\textbf{k}}_2 \times \hat{\textbf{k}}_1+5 \gamma (\textbf{e}_1 \cdot \hat{\textbf{k}}_2) \, \hat{\textbf{k}}_2 \times \textbf{e}_1}_{\text{{\tiny quadrupole}}} \\
& \underbrace{+ \gamma_3 \left\lbrace (\mathcal{B} \textbf{e}_2 + \mathcal{C} \hat{\textbf{k}}_2 ) \times \textbf{e}_1 + (\mathcal{D}  \textbf{e}_2 + \mathcal{E} \hat{\textbf{k}}_2 ) \times \hat{\textbf{k}}_1 \right\rbrace}_{\text{{\tiny octupole}}} \\
& \underbrace{-  \sum_{j} \alpha_{1j} \cos \theta_j \, \hat{\textbf{s}}_j \times \hat{\textbf{k}}_1}_{\text{{\tiny spin}}}\\
& \underbrace{- \sum_{j} K_j n_1\left( f_4(e_1) \sqrt{1-e_1^2} \frac{w_j}{2 n_1}(\hat{\textbf{s}}_j -\cos \theta_j \hat{\textbf{k}}_1 ) \right. }_{\text{{\tiny dissipative tides}}}\\
& \underbrace{\left. -f_1(e_1) \frac{w_j}{n_1} \hat{\textbf{s}}_j  + f_2(e_1)\hat{\textbf{k}}_1 +\frac{(\textbf{e}_1 \cdot \hat{\textbf{s}}_j) (6+e_1^2)}{4 (1-e_1^2)^{\frac{9}{2}}} \frac{w_j}{n_1} \textbf{e}_1 \right),}_{\text{{\tiny dissipative tides}}}
\end{aligned}
\end{equation}
\begin{equation}\label{equations_vectorielles_octupole_G2}
\begin{aligned}
\dot{\textbf{G}}_2 &= -\dot{\textbf{G}}_1 - \dot{\textbf{L}}_0 - \dot{\textbf{L}}_1,
\end{aligned}
\end{equation}
\begin{equation}\label{equations_vectorielles_octupole_e1}
\begin{aligned}
\dot{\textbf{e}}_1=&\underbrace{-\gamma \frac{(1-e_1^2)}{G_1}\left(\cos i \, \hat{\textbf{k}}_2 \times \textbf{e}_1 -2 \, \hat{\textbf{k}}_1\times \textbf{e}_1 -5 
(\textbf{e}_1 \cdot \hat{\textbf{k}}_2) \, \hat{\textbf{k}}_2\times \hat{\textbf{k}}_1  \right)}_{\text{{\tiny quadrupole}}}  \\
& \underbrace{+\frac{\gamma_3}{G_1} \left\lbrace (1-e_1^2)(\mathcal{A} \textbf{e}_1 + \mathcal{B} \textbf{e}_2 + \mathcal{C} \hat{\textbf{k}}_2 ) \times \hat{\textbf{k}}_1 + (\mathcal{D} \textbf{e}_2 + \mathcal{E} \hat{\textbf{k}}_2 ) \times \textbf{e}_1 \right\rbrace }_{\text{{\tiny octupole}}}\\
& \underbrace{-\sum_{j=0}^{1}\frac{\alpha_{1j}}{G_1}\left(\cos \theta_j \, \hat{\textbf{s}}_j \times \textbf{e}_1 +\frac{1}{2}(1-5 \cos ^2 \theta_j)\,  \hat{\textbf{k}}_1 \times \textbf{e}_1 \right)}_{\text{{\tiny spin}}} \\
& \underbrace{+ \frac{3 \mu_1 n_1}{c^2 a_1(1-e_1^2)} \, \hat{\textbf{k}}_1 \times \textbf{e}_1}_{\text{{\tiny GR}}} + \underbrace{\sum_{j=0}^{1} \frac{15}{2} 
k_{2_j} n_1 \frac{m_{(1-j)}}{m_j} \left(\frac{R_j}{a_1}\right)^5 f_4(e_1) \, \hat{\textbf{k}}_1 \times \textbf{e}_1 }_{\text{{\tiny conservative tides}}}\\
&  \underbrace{-\sum_{j=0}^{1} \frac{K_j}{\beta_1 a_1^2}\left\lbrace f_4(e_1) \frac{w_j}{2 n_1} ( \textbf{e}_1 \cdot \hat{\textbf{s}}_j )\hat{\textbf{k}}_1 \right. }_{\text{{\tiny dissipative tides}}}  \\
&  \underbrace{ \left. -\left(\frac{11}{2} f_4(e_1) \cos \theta_j \frac{w_j}{n_1} -9 f_5(e_1) \right)\textbf{e}_1 \right\rbrace, }_{\text{{\tiny dissipative tides}}}\\
\end{aligned}
\end{equation}

\begin{equation}\label{equations_vectorielles_octupole_e2}
\begin{aligned}
\dot{\textbf{e}}_2= &\underbrace{- \frac{\gamma}{G_2} \left\lbrace (1-e_1^2)\cos i \hat{\textbf{k}}_1 \times \textbf{e}_2 - 5(\textbf{e}_1 \cdot \hat{\textbf{k}}_2) \textbf{e}_1 \times \textbf{e}_2 \right. }_{\text{{\tiny quadrupole}}} \\
&\underbrace{ \left. + \frac{1}{2} \left( 1-6e_1^2 -5(1-e_1^2) \cos ^2 i + 25 (\textbf{e}_1 \cdot \hat{\textbf{k}}_2)^2 \right) \hat{\textbf{k}}_2 \times \textbf{e}_2 \right\rbrace}_{\text{{\tiny quadrupole}}} \\
& +\underbrace{\frac{\gamma_3}{G_1} \left\lbrace (\mathcal{F} + \mathcal{C} (\textbf{e}_1 \cdot \hat{\textbf{k}}_2 ) + \mathcal{E} \cos i ) \textbf{e}_2 \times \hat{\textbf{e}}_2 \right.}_{\text{{\tiny octupole}}}  \\
& + \underbrace{\left. (1-e_2^2) (\mathcal{B} \textbf{e}_1 + \mathcal{D} \hat{\textbf{k}}_1 ) \times \hat{\textbf{k}}_2 + \right. }_{\text{\tiny octupole}}\\
& \underbrace{ \left. (\mathcal{C} \textbf{e}_1 + \mathcal{E} \hat{\textbf{k}}_1 ) \times \textbf{e}_2  \right\rbrace, }_{\text{\tiny octupole}}\\
\end{aligned}
\end{equation}
\begin{equation}\label{equations_vectorielles_octupole_Lj}
\begin{aligned}
\dot{\textbf{L}}_j&= \underbrace{-\alpha_{1j}\cos \theta_j \, \hat{\textbf{k}}_1 \times \hat{\textbf{s}}_j}_{\text{{\tiny spin}}} \underbrace{-1.5 \cdot 10^{-14} \, C_j \, w_j^3 \, \hat{\textbf{s}_j}}_{\text{\tiny Spin-down (magnetic braking)}}  \\
& + \underbrace{K_j n_1\left( f_4(e_1) \sqrt{1-e_1^2} \frac{\omega_j}{2 n_1}(\hat{\textbf{s}}_j -\cos \theta_j \hat{\textbf{k}}_1 )-f_1(e_1) \frac{w_j}{n_1} \hat{\textbf{s}}_j \right. }_{\text{{\tiny dissipative tides}}} \\
& \underbrace{\left. +f_2(e_1)\hat{\textbf{k}}_1 +\frac{(\textbf{e}_1 \cdot \hat{\textbf{s}}_j) (6+e_1^2)}{4 (1-e_1^2)^{\frac{9}{2}}} \frac{w_j}{n_1} \textbf{e}_1 \right),}_{\text{{\tiny dissipative tides}}}
\end{aligned}
\end{equation}
\noindent where $\cos \theta_j = \hat{\textbf{s}}_j \cdot \hat{\textbf{k}}_1$, $\cos i=\hat{\textbf{k}}_1 \cdot \hat{\textbf{k}}_2$, and $G_j$ is the norm of $\textbf{G}_j$. We also have 
\begin{eqnarray}
n_1&=&\sqrt{\frac{\mu_1}{a_1^3}}, \\
\alpha_{1j} &=& \frac{3Gm_0 m_1 J_{2_j} R_j^2}{2a_1^3(1-e_1^2)^{(3/2)}}, \\
\gamma&=&\frac{3Gm_2 \beta_1 a_1^2}{4 a_2^3 (1-e_2^2)^{(3/2)}}, \\
K_j&=&\Delta t_j \frac{3k_{2_j} G m^2_{1-j} R_j^5 }{a_1^6}.
\end{eqnarray}
The parameters related to the octupole part of the equations are the following: 
\begin{eqnarray}
\gamma_3 &=& \frac{15Gm_2\beta_1 a_1^3(m_0-m_1)}{64 a_2^4(1-e_2^2)^{\frac{5}{2}}(m_0+m_1)}, \\
\mathcal{A} &=& 16(\textbf{e}_1 \cdot \textbf{e}_2 ), \\
\mathcal{B} &=& - \left( 1-5 (1-e_1^2) \cos ^2 i + 35 (\textbf{e}_1 \cdot \hat{\textbf{k}}_2)^2 - 8e_1^2 \right), \\
\mathcal{C} &=& 10(1-e_1^2)( \hat{\textbf{k}}_1 \cdot \textbf{e}_2 ) \cos i - 70 (\textbf{e}_1 \cdot \hat{\textbf{k}}_2 )(\textbf{e}_1 \cdot \textbf{e}_2), \\
\mathcal{D} &=& 10(1-e_1^2)( \textbf{e}_1 \cdot \hat{\textbf{k}}_2 ) \cos i, \\
\mathcal{E} &=& 10(1-e_1^2)\left[(\textbf{e}_1 \cdot \textbf{e}_2 )\cos i + (\textbf{e}_1 \cdot \hat{\textbf{k}}_2 )(\hat{\textbf{k}}_1 \cdot \textbf{e}_2) \right],  \\ \mathcal{F} &=& 5 \left[ \mathcal{B} (\textbf{e}_1 \cdot \textbf{e}_2) + \mathcal{D}(\hat{\textbf{k}}_1 \cdot \textbf{e}_2) \right].
\end{eqnarray}

\normalsize


\bsp	
\label{lastpage}
\end{document}